\documentclass[11pt]{cernrep}
\usepackage{graphicx}
\usepackage{here}
\newcommand{\bp}{{\bf b}_\perp}
\newcommand{\bT}{{\bf b}_\perp}
\newcommand{\yT}{{\bf y}_\perp}
\newcommand{\oT}{{\bf 0}_\perp}

\newcommand{\kT}{{\bf k}_\perp}
\newcommand{\sT}{{\bf s}_\perp}
\newcommand{\Dp}{{\bf \Delta}_\perp}
\newcommand{\Ds}{{\bf \Delta}_\perp^2}
\newcommand{\bi}{\begin{itemize}}
\newcommand{\ei}{\end{itemize}}
\newcommand{\be}{\begin{eqnarray}}
\newcommand{\ee}{\end{eqnarray}}
\newcommand{\bea}{\begin{eqnarray}}
\newcommand{\eea}{\end{eqnarray}}
\newcommand{\ipd}{q(x,{\bf b}_\perp)}
\begin{document}
 \title{CHROMODYNAMIC LENSING AND $\perp$ SINGLE SPIN ASYMMETRIES}
\author{M. Burkardt}
\institute{Department of Physics, New Mexico State University,
Las Cruces, NM 88003, U.S.A.}
\maketitle
\begin{abstract}
The physical interpretation of generalized parton 
distributions (in the limit $\xi=0$) as Fourier transforms
of impact parameter dependent parton distributions is discussed. 
Particular emphasis is put on the role of the target polarization.
For transversely polarized targets we expect a significant deviation
from axial symmetry for the distribution in the transverse plane.
We conjecture that this transverse distortion, in combination with
the final state provides a natural explanation for the sign of the
Sivers contribution to semi-inclusive single-spin asymmetries.
\end{abstract}

\section{INTRODUCTION}
For many years, deep inelastic scattering (DIS) experiments have 
been a useful tool for exploring hadron structure. In the Bjorken
limit, these experiments probe the parton distributions
$q(x)$ which can be defined as the expectation value of
a light-like correlation function
\be
q(x) = \int \frac{dx^-}{2\pi} \;
\langle p|\overline{q}\left(-\frac{x^-}{2},{
\bf 0_\perp}\right) \gamma^+ 
q\left(\frac{x^-}{2},{\bf 0_\perp}\right)
|p\rangle \;e^{ix^-x_{Bj}P^+}.
\label{eq:q}
\ee
Throughout this paper, we use light-cone variables
$x^\pm = \frac{1}{\sqrt{2}}\left(x^0 \pm x^3\right)$.

The physical interpretation of $q(x)$ is that of a light-cone 
momentum density of quarks in the target, where $x$ is
the fraction of the target's light-cone momentum that is
carried by the active quark. However, Eq. 
(\ref{eq:q}) provides no information about the position of the
quarks.

A generalization of Eq. (\ref{eq:q}) to non-forward
matrixelements yields the generalized parton distributions
(GPDs)\cite{1,2,3}\footnote{For a recent comprehensive review see Ref. \cite{4}.}
\be
GPD(x,\xi,t) \equiv \int \frac{dx^-}{2\pi} \;
\langle p^\prime |\overline{q}\left(-\frac{x^-}{2},{
\bf 0_\perp}\right) \gamma^+ 
q\left(\frac{x^-}{2},{ \bf 0_\perp}\right)
|p\rangle \;e^{ix^-xP^+}
\ee
with $\Delta = p-p^\prime$, $t=\Delta^2$, and $\xi (p^++
{p^+}^\prime) = -2 \Delta^+$. 
Experimentally, these GPDs can for example be probed in deeply 
virtual Compton scattering.

Recently, people became very interested in GPDs after it became 
clear that they can be linked to a number of other observables.
For example, upon integration over $x$ they can be related to
form factors. In that sense they provide a decomposition of
form factors w.r.t. the (average) light-cone momentum of the
active quark. Such information can for example be very useful
to understand the mechanism for form factors at high momentum
transfer. Another application of GPDs is that knowing GPDs would
enable us 
to determine a quantity that can be identified with the
total (spin+orbital) angular momentum carried by the quarks in
the nucleon. If fact, it becomes more and more clear that GPDs 
could provide us with key information about the orbital 
angular momentum structure of the nucleon.

However, there is another, very interesting, piece of
information about the structure of hadrons that GPDs can provide,
namely they can teach us how partons are distributed in the
transverse plane. Discussing this connection and possible
consequences will be the main purpose of
this talk.

\section{IMPACT PARAMETER DEPENDENT PARTON DISTRIBUTIONS}

In order to help us understand, in simple physical pictures,
the kind of information that is contained in GPDs, we will in the
following explore the analogy to form factors. Indeed, we can 
write the definition of the GPDs $H$ and $E$ in a form that
emphasizes this analogy
\be
\left\langle p^\prime \left|
\hat{O}
\right|p\right\rangle
=H(x,\xi,\Delta^2)\bar{u}(p^\prime)\gamma^+ u(p)
+ E(x,\xi,\Delta^2)\bar{u}(p^\prime)
\frac{i\sigma^{+\nu}\Delta_\nu}{2M} u(p)
\label{eq:defHE}
\ee
with $\hat{O} \equiv \int \frac{dx^-}{2\pi}e^{ix^-\bar{p}^+x}
\bar{q}\left(-\frac{x^-}{2}\right)
\gamma^+ q\left(\frac{x^-}{2}\right)
$. The only difference between Eq. (\ref{eq:defHE}) and the
definition of the Dirac and Pauli form factors $F_1$ and $F_2$ is
the fact that the current density operator is substituted by
the operator $\hat{O}$. When sandwiched between momentum
eigenstates, this operator $\hat{O}$ acts like a ``filter'' that lets
through only quarks that carry a certain momentum fraction $x$
and GPDs are the form factors of this momentum filter.

The forward matrix element of the vector current gives the
charge. Form factors are the non-forward matrix element of the 
vector current operator. By taking the Fourier transform of the
form factor we can learn how the charge (i.e. the physical
quantity that is related to the forward matrix element) is 
distributed in position space.

\begin{figure}
\unitlength1.cm
\begin{picture}(9,4)(-2.3,1)
\put(1.4,4.25){operator}
\put(1.6,2.8){$\bar{q}\gamma^+q$}
\put(-0.2,1.6){
$\int\! \frac{dx^- e^{ixp^+x^-}\!\!\!\!\!}{4\pi}\,
\bar{q}\!\left(\!\frac{-x^-}{2}\!\right)\gamma^+
q\!\left(\!\frac{x^-}{2}\!\right)$}
\put(4.5,5){\line(0,-1){4}}
\put(7,5){\line(0,-1){4}}
\put(9.5,5){\line(0,-1){4}}
\put(4.7,4.5){forward}
\put(4.7,4){matrix elem.}
\put(5.7,2.8){$Q$}
\put(5.5,1.6){$q(x)$}
\put(7.2,4.5){off-forward}
\put(7.2,4){matrix elem.}
\put(7.9,2.8){$F(t)$}
\put(7.5,1.6){$H(x,0,t)$}
\put(9.7,4.25){position space}
\put(10.5,2.8){$\rho({\vec r})$}
\put(10.2,1.6){{$q(x,{\bf b}_\perp)$}}
\put(-0.2,3.7){\line(1,0){12}}
\end{picture}
\label{fig1}
\caption{Illustration of the analogy between the {\sl form factor} 
$\leftrightarrow$ {\sl charge distribution} correspondence and the 
{\sl GPD} $\leftrightarrow$ {\sl impact parameter dependent parton
distribution} correspondence.}
\end{figure}

In the case of GPDs, the forward matrix element gives the usual
parton distribution functions (PDFs). By analogy with the form factor
of the vector current one would therefore expect that some Fourier
transform of GPDs provides information about how the usual PDFs are 
distributed in position space (Fig. \ref{fig1}). 
Working out the details of this
 will be the subject of the first part of these notes.

Of course, since the usual PDFs already measure the longitudinal
momentum of the quarks, Heisenberg's uncertainty principle allows us
to measure only the transverse position of the partons. Because of
that we will in the following only consider the case where the
momentum transfer in GPDs is purely transverse (i.e. 
$\xi\propto\Delta^+=0$).\footnote{Note that if one makes only an
approximate measurement of the longitudinal momentum then
one can still make an (approximate) measurement of the longitudinal
position, as long as the Heisenberg inequality is obeyed \cite{ji}.}

Before we can proceed and derive the connection between GPDs and
PDFs in transverse position (``impact parameter'') space, we need
to define what we mean by impact parameter dependent PDFs. For this
purpose we introduce wave packets that have a sharp longituninal
momentum and that are localized in transverse position \cite{5,6,7}
\be
\left| p^+, {\bf R}_\perp = {\bf 0}_\perp,\lambda\right\rangle
\equiv {\cal N}
, \label{eq:local}\int d^2{\bf p}_\perp 
\left|p^+,{\bf p}_\perp,\lambda \right\rangle
\ee
where ${\cal N}$ is a normalization constant, such that
$(2\pi)^2 \int d^2{\bf p}_\perp \left|{\cal N}\right|^2=1$. 
This state is
localized in impact parameter space in the sense that it is an
eigenstate of the $\perp$ center of (longitudinal) momentum
\be
{\bf R}_\perp \equiv \frac{1}{p^+} 
\int dx^-d^2{\bf x}_\perp\, T^{++}(x) {\bf x}_\perp
,
\ee
where $T^{++}$ is the component of the energy momentum tensor
that describes the light-cone momentum density.
The parton representation for the $\perp$ center of momentum is the 
weighted average of $\perp$ parton positions, where the
weight factors are the fractions of $p^+$ momentum carried by each 
parton, i.e. ${\bf R}_\perp= \sum_i x_i {\bf r}_{\perp i}$.
Working with this transversely localized state is in many ways 
analogous to working in the center of mass frame in nonrelativistic
systems.

Using this state, we can now define what we mean by {\bf impact
parameter dependent parton distributions}. For example for the
unpolarized distributions, we define \cite{8}
\be
q(x,{\bf b}_\perp) \equiv\! 
\int \!\frac{dx^-\!\!\!}{4\pi\!} 
\left\langle p^+\!,{\bf 0}_\perp \right|
\bar{q}\left(-\frac{x^-\!\!}{2}\!,{\bf b}_\perp\right)
\gamma^+ q\left(\frac{x^-\!\!}{2}\!,{\bf b}_\perp\right)
\left|p^+\!,{\bf 0}_\perp\right\rangle 
e^{ixp^+x^-}.
\label{ipdpd}
\ee
In gauges other than light-cone gauge a straight line gauge string
needs to be included in Eq. (\ref{ipdpd}). A very similar definition
can be given for the polarized impact parameter dependent parton 
distribution $\Delta q(x,{\bf b}_\perp)$.

Using translation invariance it is straightforward to relate
$\ipd $ to GPDs
\bea
q(x,{\bf b}_\perp)&\equiv&\int \!\!dx^-
\left\langle p^+\!,{\bf R}_\perp = {\bf 0}_\perp
\right|
\bar{q}(-\frac{x^-\!}{2}\!,{\bf b}_\perp)
\gamma^+ q(\frac{x^-\!}{2}\!,{\bf b}_\perp)
\left|p^+\!,{\bf R}_\perp={\bf 0}_\perp
\right\rangle e^{ixp^+x^-}
\label{qH} \\
&= &\left|{\cal N}\right|^2
\int\!\!\!d^2 {\bf p}_\perp\! \!\int \!\!\!d^2 
{\bf p}_\perp^\prime \!\!\int\! \!dx^-\!
\left\langle p^+\!,{\bf p}_\perp^\prime \right|
\bar{q}(-\frac{x^-\!}{2}\!,{\bf b}_\perp)
\gamma^+ 
q(\frac{x^-\!}{2}\!,{\bf b}_\perp)\left|p^+\!,
{\bf p}_\perp \right\rangle 
e^{ixp^+x^-}
\nonumber\\
&=&\left|{\cal N}\right|^2
\int \!\!\!d^2 {\bf p}_\perp\! \!\int \!\!\!d^2 
{\bf p}_\perp^\prime \!{ \int\! \!dx^-
\!
\left\langle p^+\!,{\bf p}_\perp^\prime \right|
\bar{q}(-\frac{x^-\!}{2}\!,{\bf 0}_\perp)
\gamma^+ q(\frac{x^-\!}{2}\!,{\bf 0}_\perp)\left|p^+\!,
{\bf p}_\perp \right\rangle 
e^{ixp^+x^-}}
\nonumber\\ 
& &\quad \quad \quad \quad \quad \quad
\quad\quad\quad \quad \times
e^{i{\bf b}_\perp\cdot ({\bf p}_\perp-
{\bf p}_\perp^\prime)}
\nonumber\\ 
&=& \left|{\cal N}\right|^2
\int \!\!\!d^2 {\bf p}_\perp\! \int \!\!\!d^2 
{\bf p}_\perp^\prime {
H\left(x,0,-\left({\bf p}^\prime_\perp -{\bf p}_\perp
\right)^2\right)}
e^{i{\bf b}_\perp\cdot 
({\bf p}_\perp-{\bf p}_\perp^\prime)} \nonumber
\end{eqnarray}
Upon switching variables to sums and differences of momenta one thus 
finds that the GPD $H(x,0,-\Ds)$ is the Fourier transform of 
$\ipd$ \cite{6}
\be
q(x,{\bf b}_\perp)=\int \frac{d^2{\bf \Delta}_\perp}
{(2\pi)^2}  
H(x,0,-{\bf \Delta}_\perp^2) e^{i{\bf b}_\perp \cdot
{\bf \Delta}_\perp}.
\ee
Besides being the Fourier transform of GPDs, $\ipd$ satisfies a
number of positivity constraints. For example \cite{8}
\bea
\begin{array}{ccc} \ipd >0 & \mbox{for} & x>0\\
\ipd <0 & \mbox{for} & x<0 \end{array},
\eea
where the minus sign for $x<0$ 
follows from charge conjugation. The proof of
these positivity constraints parallels the proof that the usual
PDFs are positive. As a result, one can also derive various
``Soffer-type'' inequalities among  PDFs in impact parameter 
space \cite{9}.

For the practitioner, positivity constraints are useful because they
provide model-independent theoretical constraints on any 
phenomenological ansatz for GPDs.
However, a much more important consequence
of these inequalities is that they allow
a probabilistic interpretation for $\ipd$, which indicates that
$\ipd$ has a physical meaning above and beyond being the Fourier
transform of $H(x,0,-\Ds)$.

\subsection{Discussion}
Knowledge of GPDs for purely transverse momentum transfer allows
probing parton distributions in impact parameter space. This is
completely novel information about the nucleon structure and will
provide interesting tests for our understanding of the quark-gluon
structure of hadrons.

\begin{figure}
\unitlength1.cm
\begin{picture}(10,7.6)(3,20.5)
\includegraphics{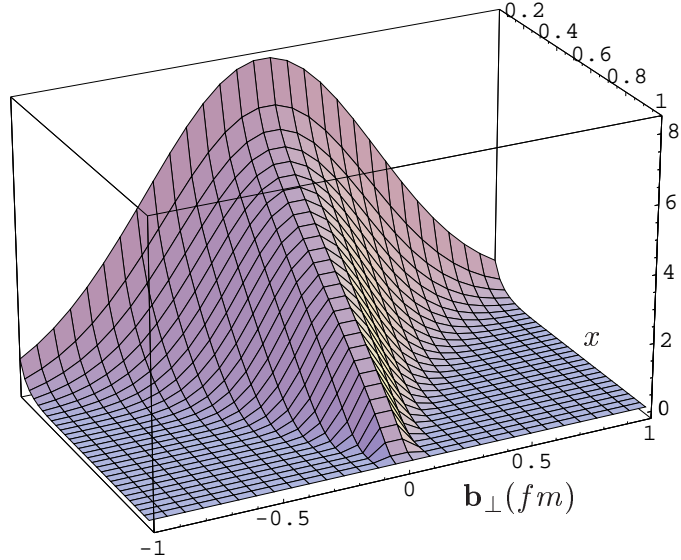}
\end{picture}
\caption{Anticipated shape of $q(x,\bp)$ (qualitative)}
\label{fig2}
\end{figure}

The reference point for the impact parameter dependent PDFs is the
$\perp$ center of momentum ${\bf R}_\perp \equiv \sum_{i=q,g}
x_i {\bf r}_{\perp,i}$. When $x\rightarrow 1$, the active quark
becomes the center of momentum and as a result the transverse width
of $b(x,\bp)$ should go to zero. Note that this does not
mean that the transverse size of the nucleon goes to zero, since for
example the distance ${\bf B}_\perp$ between the active quark and
the center of momentum of the spectators can remain finite as
$x\rightarrow 1$, since $\bp = (1-x) {\bf B}_\perp$. However, what 
the vanishing $\perp$ width implies is that $H(x,0,t)$ should become
$t$-independent as $x\rightarrow 1$.\footnote{Of course, at the same
time $H(x,0,t)$ should go to zero since $q(x)$ goes to zero
as $x\rightarrow 1$ and therefore whether or not the form factor
receives a significant contribution from $x\rightarrow 1$ 
(``Feynman mechanism'') depends on details.}. 
For decreasing $x$, one expects the size of the nucleon to grow,
because when $x\sim \frac{m_\pi}{M}$ one should see the pion
cloud and for even smaller $x$ a logarithmic growth of the
$\perp$ size with $\frac{1}{x}$ should set in \cite{weiss}.
Of course, when $x$ decreases, not only the width of $q(x,\bp)$ 
should increase but also its magnitude since there are more quarks at
small $x$.
In order to gain some intuitive understanding about this behavior, 
we have plotted $q(x,\bp)$ in Fig. \ref{fig2} for
a simple model that has all these features built in.

Note that deeply virtual Compton scattering (DVCS) experiments probe
always $\Delta^+=0$. Therefore from the point of view of DVCS, the
limit $\Delta^+=0$ is unphysical and can only be reached by 
extrapolation. However, this task is facilitated by the fact that
the $x$-moments of $E(x,\xi,t)$ have a polynomial dependence on
$\xi$ and therefore, at least theoretically, the extrapolation
to $\xi=0$ can be done model independently.

So far we did not discuss the scale-dependence of GPDs.
Adding scale dependence to our above results is trivial since
QCD evolution, which addresses only the divergent part of the 
$Q^2$ dependence is local in impact parameter space, i.e.
there is no mixing between different $\bp$ and there are separate
DGLAP evolution equations for each $\bp$.
This is consistent with the fact that the evolution of GPDs is
$t$-independent. Of course, this is valid only for $Q^2\gg t$ and 
therefore $1/Q^2$ limits the transverse ``pixel size'' in $q(x,\bp)$.

\section{THE PHYSICS OF $E(x,0,-\Dp^2)$}
For $\Delta^+=0$, the GPD $E(x,0,-\Dp^2)$ only contributes to helicity
flip amplitudes
\bea
\int \frac{dx^-}{4\pi} e^{ip^+x^- x}
\left\langle P\!\!+\!\!\Delta,\! \uparrow\!
\left| \bar{q}\!\left(\!\frac{-x^-}{2}\!\right)\gamma^+
q\!\left(\!\frac{x^-}{2}\!\right)
\right| P,\! \uparrow \right\rangle
&=&H(x,\!0,\!-{\bf \Delta}_\perp^2) \label{HE}
\\
\int \frac{dx^-}{4\pi} e^{ip^+x^- x}
\left\langle P\!\!+\!\!\Delta,\! 
\uparrow\!\left| \bar{q}\!\left(\!\frac{-x^-}{2}\!\right)\gamma^+
q\!\left(\!\frac{x^-}{2}\!\right)
\right| P,\!\downarrow\right\rangle
&=& -\frac{\Delta_x\!\!-\!i\Delta_y}{2M}E(x,\!0,\!-{\bf \Delta}_\perp^2).
\nonumber
\eea
Therefore, in order to understand the physics of $E(x,0,-\Dp^2)$ we 
need to consider states that are not helicity eigenstates.
The contribution from $E$ is maximal in states that have equal 
probability from both helicities and we therefore consider the state
\be
\left| X\right\rangle \equiv \frac{1}{\sqrt{2}}\left[
\left| p^+, {\bf R}_\perp = {\bf 0}_\perp,\uparrow\right\rangle +
\left| p^+, {\bf R}_\perp = {\bf 0}_\perp,\downarrow\right\rangle
\right],
\ee
which one may interpret as a state that has a transverse center of 
momentum localized at the origin and that is polarized in the
$\hat{x}$ direction in the infinite momentum frame.\footnote{The rest 
frame interpretation of the state may be subject to
Wigner-Melosh rotations, i.e. when viewed from the rest frame, this
state corresponds to a nucleon polarized in the $x$ direction plus 
some relativistic corrections.} 
We denote the {\sl unpolarized} quark distribution in impact parameter
space for this transversely polarized state by $q_X(x,\bp)$, i.e.
\be
q_X(x,\bp)\equiv
\! 
\int \!\frac{dx^-\!\!\!}{4\pi\!} 
\left\langle X \right|
\bar{q}\left(-\frac{x^-\!\!}{2}\!,{\bf b}_\perp\right)
\gamma^+ q\left(\frac{x^-\!\!}{2}\!,{\bf b}_\perp\right)
\left|X\right\rangle 
e^{ixp^+x^-}.
\ee
In order to relate $q_X(x,\bp)$ to GPDs, we follow the same steps as 
in Eq. (\ref{qH}). The only difference is that one now obtains
both matrix elements that are diagonal in the target spin as well
as matrix elements that involve a target spin flip.
Making use of Eq. (\ref{HE}) one thus finds after some integration 
by parts that the unpolarized quark distribution in a 
transversely polarized nucleon is the same as the unpolarized
quark distribution in a longitudinally polarized (or unpolarized)
nucleon plus a correction term. The correction term is proportional
to the gradient of the Fourier transform of 
$E(x,0,\!-{\bf \Delta}_\perp^2)$
\be
q_X(x,\!{\bf b_\perp}) = q(x,\!{\bf b_\perp})
-
\frac{1}{2M}\frac{\partial}{\partial b_y}\!
\int \!\!\frac{d^2\Dp}{(2\pi)^2} 
E(x,0,\!-{\bf \Delta}_\perp^2)
e^{i{\bf b}_\perp\cdot{\bf \Delta}_\perp}.
\label{qX}
\ee
If the nucleon is longitudinally polarized then rotational symmetry
around the $z$ axis implies that the impact parameter dependent
PDF $q(x,\bp)$ is axially symmetric (depends only on $\bp^2$).
However, when the nucleon is transversely polarized then there is
no reason why $q_X(x,\bp)$ should be axially symmetric.
The distortion is described by the gradient of the Fourier transform 
of $E(x,0,\!-{\bf \Delta}_\perp^2)$.

The direction of the distortion can be easily understood from a simple
classical picture: In DIS one probes the $+$ component of the current.
Since $j^+=j^0+j^z$, the distortion arises because an orbital
motion around the $x$ axis produces a $j^z$-current that is
asymmetric w.r.t. $\pm \hat{y}$. This explains from an intuitive
point of view why the $\hat{y}$ derivative appears in Eq. (\ref{qX}).

In order to understand the magnitude of the distortion, we would 
have to know the function $E(x,0,\!-{\bf \Delta}_\perp^2)$.
However, even without knowing $E(x,0,\!-{\bf \Delta}_\perp^2)$ we
can still estimate the mean effect by evaluating the
transverse flavor dipole moment that results from this distortion
\be
d^q_y \equiv \int\!\! dx\!\int \!\!d^2\bp
q_X(x,\bp) b_y
=\frac{1}{2M} 
\int     dx E_q(x,0,0) = \frac{1}{2M} F_{2q}(0) = 
\frac{\kappa_{q}^p}{2M},
\label{dq}
\ee 
where $\kappa_{q}^p$ is the anomalous magnetic moment contribution
to the proton from flavor $q$. A simple $SU(3)$ analysis
(neglecting the small strange quark contribution)\footnote{Note that
$\kappa_u^p-\kappa_d^p=\kappa^p-\kappa^n \approx 3.7$ is independent of the strange
magnetic moment.} yields
$\kappa_u^p \approx 1.67$ and $\kappa_d^p \approx -2.03$ i.e.
the resulting flavor dipole moments are on the order of $0.1-0.2\,fm$,
which is a significant effect.

\begin{figure}
\unitlength1.cm
\begin{picture}(10,10.1)(1.3,2.3)
\includegraphics{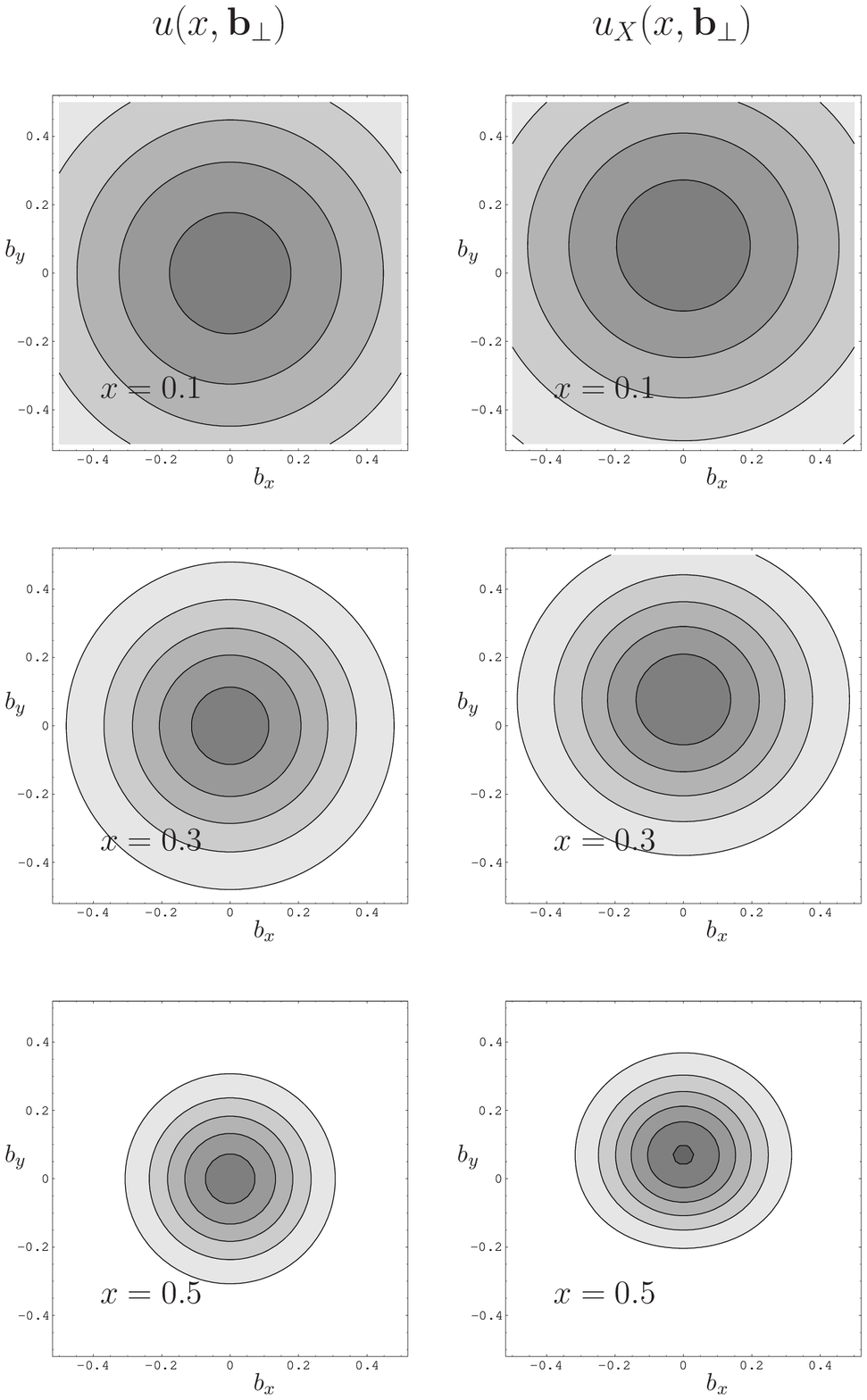}
\put(7,0){\includegraphics{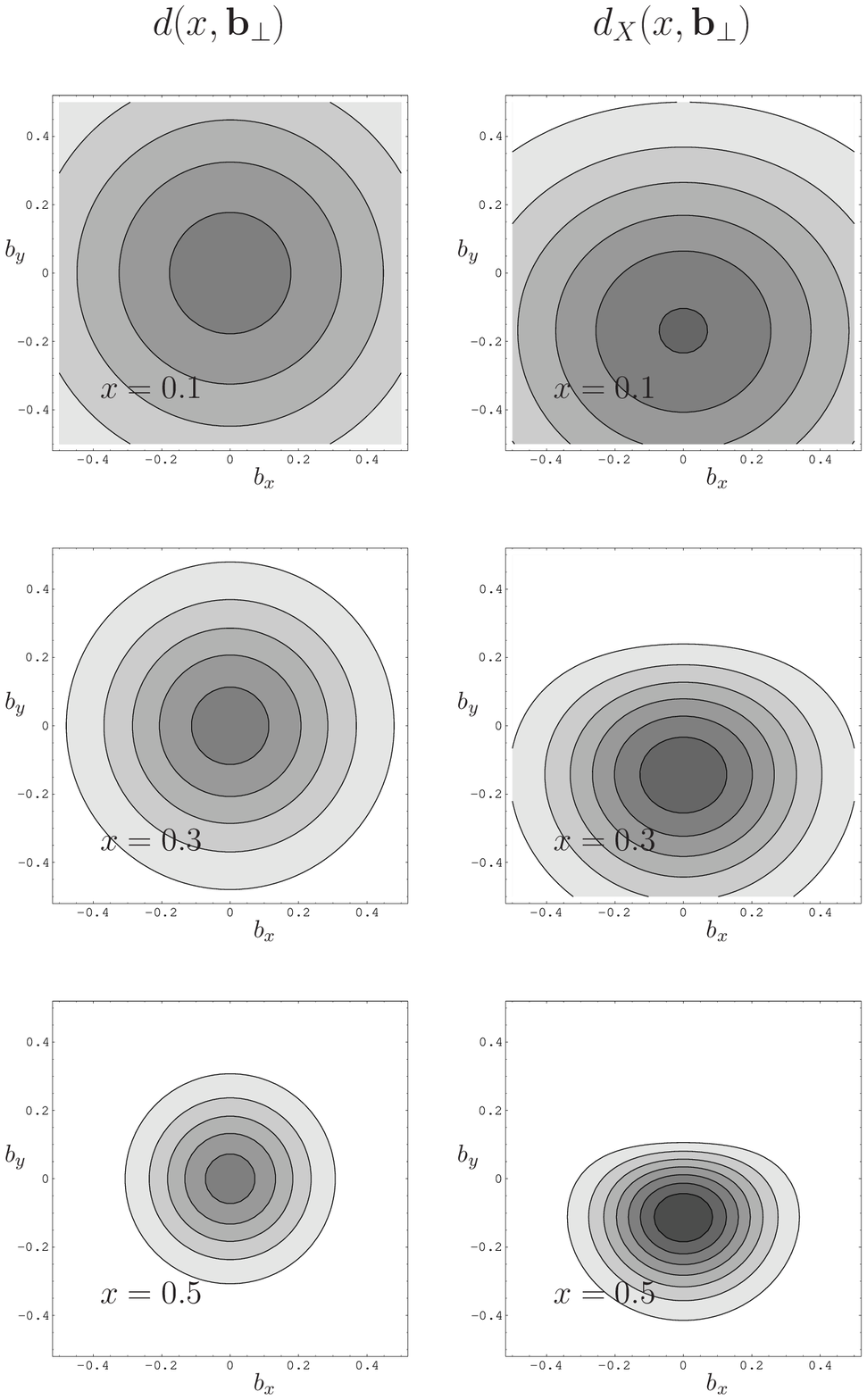}}
\end{picture}
\caption{Parton distributions in impact parameter space
for the simple model for $x=0.1, 0.3, 0.5$. For each plot the
grayscales are normalized to the central value.
$1^{st}$ column: (unpolarized) distribution of $u$ quarks in a 
longitudinally polarized nucleon. $2^{nd}$ column: same for a
nucleon that is polarized in the $\hat{x}$ direction.
$3^{rd}$ and $4^{th}$ columns: same for $d$ quarks.
}
\label{fig3}
\end{figure}

In order to illustrate the magnitude of the anticipated distortion,
we take the model for $H_q(x,0,t)$ that was used in Fig. \ref{fig1}
and as a model for $E_q(x,0,t)$ we make the ansatz
\be
E_u(x,0,t) = \frac{1}{2}\kappa_u H_u(x,0,t)
\quad \quad \quad \quad 
E_d(x,0,t) = \kappa_d H_d(x,0,t).
\ee
The factor $\frac{1}{2}$ accounts for the fact that $H_u=2H_d$ in 
this very simple model. Of course, we do not really expect that
$H$ and $E$ are proportional. But for the purpose of providing
a rough picture of the expected effects, this crude ansatz may be
useful. The resulting parton distributions in impact parameter space
are shown in Fig. \ref{fig3}.
Even though any details (e.g. $x$-dependence) of the distortion
are of course model dependent, the mean
magnitude of the effect is constrained by Eq. (\ref{dq}) and
thus model-independent. Fig. \ref{fig3} thus clearly illustrates 
that the anticipated distortion is quite significant.
Notice that the opposite signs of the distortion for $u$ and $d$
quarks are due to the fact that $\kappa_u$ and $\kappa_d$ have
opposite signs. The fact that the distortion is larger for
$d$ than for $u$ quarks is due to the fact that $E_u$ and $E_d$
have about the same magnitude, but $H_u$ is twice as large as $H_d$.

\section{$\perp$ SINGLE SPIN ASYMMETRIES}
In the previous section we demonstrated that quark distribution
functions in a transversely polarized nucleon are expected to have
a significant left-right (w.r.t. the spin) asymmetry in impact
parameter space. In this section we would like to add some 
speculations about possible ramifications of this effect for other
experiments. In particular, we will focus on the transverse
single spin asymmetry in semi-inclusive photo-production of mesons
off a transversely polarized target.

For example, let us consider the inclusive production of $\pi^+$
and $\pi^{0}$ mesons off nucleons that are polarized into the plane,
with unpolarized photons coming in from the $-\hat{z}$ direction.
Since $e_u^2=4e_d^2$, and since $u\rightarrow \pi^+,\pi^0$ 
fragmentation is `favored', most $\pi^+,\pi^0$
mesons result from an initial up quark that has been knocked out.
At the quark level, several mechanisms have been proposed that
can give rise to a left-right asymmetry (relative to the nucleon spin)
of the produced mesons. In the Collins effect, a transversely 
polarized quarks fragments with a left-right asymmetry into mesons.
Here we are not discussing this effect. Instead we discuss the
Sivers effect where the outgoing $u$ quark has already a
left-right asymmetry {\it before} it fragments. Of course, a 
left-right asymmetry in the momentum ${\bf k_q}$ of a quark in the 
nucleon is inconsistent with time-reversal invariance
[${\bf k_q}\cdot ({\bf S}_N\times {\bf p_N})$ is T-odd] and therefore
any ${\bf k_q}$ asymmetry can only arise from the final state 
interactions (FSI) of the struck quark as it escapes from the target.
The FSI can be conveniently included in a gauge invariant
definition of unintegrated parton densities \cite{13,10}
\be
P(x,\kT,\sT) &=& \int \frac{dy^- d^2\yT }{16\pi^3}
e^{-ixp^+y^-+i\kT \cdot \yT } \label{eq:P}
\left\langle p \left|\bar{q}(0,y^-,\yT) W^\dagger_{y\infty}
\gamma^+ W_{0\infty} q(0)\right|p\right\rangle .
\label{P}
\ee
$W_{y\infty}=P\exp \left(-ig\int_{y^-}^\infty dz^- A^+(y^+,z^-,\yT)
\right)$ indicates a path ordered Wilson-line operator 
going out from the point $y$ to infinity.
Starting from Eq. (\ref{P}) one finds for the mean transverse momentum
\cite{11}
\be
\langle {\bf k}_x\rangle  = \int_0^\infty dy^-
\left\langle p,s \left|\bar{q}(0,0^-,\oT) W^\dagger_{0y}
\gamma^+ G^{+x}(y^-,\oT) W_{0y}
q(0,0^-,\oT)\right|p,s\right\rangle. \label{k}
\ee
where $G^{\mu \nu}$ is the QCD field strength tensor.
Apart from the gauge link factors $W_{0y}$, which are only there
to make Eq. (\ref{k}) gauge invariant, this result has a very simple
physical interpretation: The mean transverse momentum of the
outgoing quark is obtained by integrating the transverse force
(from $G^{+x}$) along its outward path 
(which is along the light-cone for a high-energy process).
Although Eq. (\ref{k}) (without the gauge links) has been written 
down a long time ago \cite{12}, the momentum space expression has not
helped much to estimate the size, or even the sign, of 
$\langle k_x\rangle $ in QCD. 
As an application of the impact parameter
picture, we will attempt in the following to predict the sign of
the Sivers asymmetry.

For this purpose, we first make use of Galilei invariance under 
$\perp$ boosts to rewrite Eq. (\ref{k}) in impact parameter space
\cite{14}
\be
\langle {\bf k}_x \rangle &=& \int_0^\infty \!\!\!dy^-\int d^2\bT
\left\langle 
 p^+, {\bf R}_\perp = {\bf 0}_\perp, s
\left|\bar{q}(0,0^-,\bT) W^\dagger_{0^-\bT,y^-\bT}
\gamma^+ \right.\right. \label{kb}\\
& &\quad\quad\quad\quad\quad\quad\quad\quad\quad\quad\quad\quad
\times
\left. \left. G^{+x}(y^-,\bT) W_{0^-\bT,y^-\bT}
q(0,0^-,\bT)\right|
 p^+, {\bf R}_\perp = {\bf 0}_\perp,s
\right\rangle. \nonumber
\ee
The r.h.s. of Eq. (\ref{kb}) can be interpreted as the correlation
between the transverse position of the quark and the
transverse impulse that the quark experiences from the FSI, 
when it is knocked out from that transverse position.
Intuitively, we would expect the FSI on average to be attractive, 
since it costs energy to build up the `string' of gauge fields
that connects the escaping quark with the spectators before 
quark pair creation leads to a breaking of this `string'.
Although the actual force that acts on the struck quark is a 
complicated superposition of forces from all the spectators, we still
expect that the average force still has some of the features of this
semi-classical string picture and hence we expect (on average) an
attractive force on the outgoing quark. We should emphasize that
many phenomenological models \cite{feng,dsh} have this feature implicitly
built in.
\begin{figure}
\unitlength1.cm
\begin{picture}(10,2.5)(2,19)
\includegraphics{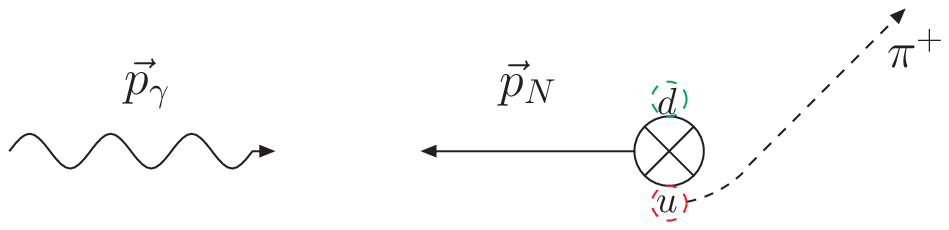}
\end{picture}
\caption{$\gamma p$ collision in the Breit frame and 
anticipated asymmetry for inclusive $\pi^+$ production from
protons that are polarized into the plane.}
\label{fig4}
\end{figure}

Consider now a photon, which is moving in the $-\hat{z}$ direction,
that collides with a nucleon that is polarized in the $+\hat{y}$
direction. According to the results from the previous section,
when viewed from the Breit frame,
the $u$ quarks tend to be displaced in the $-\hat{x}$ direction
in impact parameter space. If, as we argued above, there is on average
an attractive force on the $u$ quark after it has been struck by the 
photon, then that Force should have a component in the $+\hat{x}$
direction. We therefore expect that $\langle k_x \rangle >0$.
If we reverse the nucleon spin then the distortion in transverse
position space gets reversed and $\langle k_x \rangle $ changes sign, as it should be. Explicit model calculations \cite{feng,dsh}
confirm these results.
However, we should emphasize that our results
for the Sivers asymmetry are model independent in the sense that
we do not specify details of the FSI --- we only postulate that they
are on average attractive (towards the spectators).

Another model independent result that we have derived is that the sign
of the Sivers asymmetry is essentially\footnote{Here one assumes only
that these functions don't have fluctuating signs.} 
determined if one knows the sign of the anomalous magnetic moments
contribution from a given quark flavor and the sign 
(attractive or repulsive) of the FSI. A similar correlation has been
observed in Ref. \cite{13}. As a result we expect for example that
the Sivers adymmetries for $u$ and $d$ quarks have opposite signs.
\section*{ACKNOWLEDGEMENTS}

I would like to thank X. Ji and N. Makins for very stimulating
discussions. This work was supported in part by the DOE under
grant number DE-FG03-95ER40965.

\end{document}